\documentstyle[12pt]{article}

\def\hybrid{\topmargin -20pt  \oddsidemargin 0pt
      \headheight 0pt   \headsep 0pt
      \textwidth 6.25in 
      \textheight 9.5in 
      \marginparwidth .875in
      \parskip 5pt plus 1pt   \jot = 1.5ex}

\hybrid
\begin{document}
\def\beq{\begin{equation}}
\def\eeq{\end{equation}}
\def\beqa{\begin{eqnarray}}
\def\eeqa{\end{eqnarray}}
\def\beq{\begin{equation}}
\def\eeq{\end{equation}}
\def\beqa{\begin{eqnarray}}
\def\eeqa{\end{eqnarray}}
\sloppy
\renewcommand{\arraystretch}{1.6}
\newcommand{\be}{\begin{equation}}
\newcommand{\eq}{\end{equation}}
\begin{titlepage}
\begin{center}
\hfill HUB-EP-97/05\\
\hfill hep-th/9702010\\
\vskip .6in
{\bf STATIC $N=2$ BLACK HOLES FOR QUADRATIC PREPOTENTIALS}
\vskip .2in
{K. Behrndt\footnote{behrndt@qft2.physik.hu-berlin.de} and 
W. A. Sabra\footnote{sabra@qft2.physik.hu-berlin.de}}\hfill
\vskip 0.5cm
\hfill

{\em Humboldt-Universit\"at zu Berlin,\\
Institut f\"ur Physik,\\ 
D-10115 Berlin, Germany}\\
\end{center}
\vskip .5in
\begin{center} {\bf ABSTRACT } 
\end{center}
\begin{quotation}\noindent
We employ the principle of minimal central charge and study the entropy
of $N=2$ black holes corresponding to the most general quadratic prepotential.
We also give a static black hole solution for these models in which the 
scalar moduli are non constants. Finally, we speculate on the microscopic
origin for our solution.
\end{quotation}
\end{titlepage}
\vfill
\eject
\newpage
\newcommand{\NP}[3]{{ Nucl. Phys.} {\bf B #1} {(19#2)} {#3}}
\newcommand{\PL}[3]{{ Phys. Lett.} {\bf #1} {(19#2)} {#3}}
\newcommand{\PRL}[3]{{ Phys. Rev. Lett.} {\bf #1} {(19#2)} {#3}}
\newcommand{\PR}[3]{{ Phys. Rev.} {\bf #1} {(19#2)} {#3}}
\newcommand{\IJ}[3]{{ Int. Jour. Mod. Phys.} {\bf #1} {(19#2)}
  {#3}}
\newcommand{\CMP}[3]{{ Comm. Math. Phys.} {\bf #1} {(19#2)} {#3}}
\newcommand{\PRp} [3]{{ Phys. Rep.} {\bf #1} {(19#2)} {#3}}
\newcommand{\MPLA} [3] { {Mod. Phys. Lett} {\bf A #1} {(19#2)} {#3}}
\renewcommand{\thefootnote}{\alph{footnote}}
After an intensive study of $N=4$ black holes, there has been recently a 
considerable progress in the understanding of $N=2$ black hole in
four dimensional space-time \cite{fe/ka}-\cite{sa}. For $N=2$, however, 
one expects quantum 
corrections both at the perturbative and the 
nonperturbative level. The principle of minimal central charge
derived using the underlying special geometry structure of the moduli 
space of the scalars of the vector 
supermultiplets, is the statement that the central charge 
of the $N=2$ supersymmetry is extremum at 
the horizon of the black hole and gives the area of the horizon. 
Moreover, the horizon acts as an attractor for the scalar fields. 
Regardless of the values of the moduli at spatial infinity, 
the scalar fields at the horizon are unique and solely depend on  
conserved electric and magnetic charges. In $N=2$ models resulting from 
string compactifactions on Calabi-Yau threefolds, the fixed values of 
the moduli at the horizon and hence 
the entropy depend on the topological data 
of the compactified space. The charges correspond to the gauge group 
$U(1)^{n+1},$ 
where $n$ is the number
of vector multiplets and the extra $U(1)$ factor is due to the graviphoton. 
The special geometry \cite{CDFP} of the couplings of vector multiplets in 
$N=2$ 
supergravity is completely determined by the symplectic 
section $(L^I , M_I)$, which obey the symplectic constraint
\be
i(\bar L^IM_I-L^I\bar M_I)=1 \eq 
and  depend on the complex scalar fields of the
vector multiplets $z^{A}$ ($A=1,\cdots,n$) parameterizing a special
K\"ahler manifold with the metric $g_{A\bar{B}}=\partial_{A}
\partial_{\bar{B}} K(z,\bar{z})$, where $K(z,\bar{z})$ is the K\"ahler
potential and is given by
\be 
e^{-K} =
i (\bar{X}^I F_I - X^I \bar{F}_{I}), \qquad L^I=e^{K\over2}X^I,\quad 
M_I=e^{K\over2} F_I.
\eq 

The combined set of equations of motion and
Bianchi identities is invariant under the symplectic transformation
$\Omega \in Sp(2(n+1))$
\be
\pmatrix{L^I\cr M_I}\rightarrow \Omega\pmatrix{L^I\cr M_I}.
\eq

The relation between $M_I$ and $L^I$ depends on the embedding of the 
isometry group of the special K\"ahler manifold into the symplectic 
group \cite{sym,yaya}. For certain embeddings, a holomorphic 
prepotential $F$ exists, for which $F_I=\partial_IF$. In $N=2$ supersymmetric 
theories, whether a holomorphic prepotential 
exists 
or not, the central charge and the 
$N=2$ BPS spectrum can be expressed in terms of the symplectic section 
by \cite{CDFP}
\be
M^2_{BPS}=|Z|^2 = |q_I L^I - p^I M_I|^2,
\eq
where the $q_I$ and $p^I$ are the electric and magnetic charges 
respectively. 
The black hole ADM mass is given by the central charge of the $N=2$ 
supersymmetry algebra taken at spatial 
infinity 
\be
M^2_{ADM}=\vert Z\vert_\infty^2=\vert Z(z_\infty^A, \bar z_\infty^A)\vert^2.
\eq
Here $z^A_\infty$ $(A=1,\cdots,n),$ are the values of 
the moduli at spatial infinity.

The black hole metric is asymptotically flat at infinity and near the 
horizon the metric describes a Bertotti-Robinson universe. In both limits,
the full $N=2$ supersmmetry is preserved and the interpolating fields 
configuration between these two maximally 
supersymmetric field configurations breaks half of the 
supersymmetry, meaning that we are dealing with BPS states.

Assuming that the moduli fields take constant values from the horizon to 
spatial infinity, we 
obtain the so called double extreme black holes, in which case
$M^2_{BR}=M^2_{ADM}$ and the metric is that of Reissner-Nordstr{\o}m.

In addition to four sporadic cases, homogeneous special K\"ahler 
manifolds correspond to the moduli space 
${SU(1,1)\over U(1)}\times{SO(2,n)\over SO(2)\times SO(n)}$ and  
$SU(1,n)\over U(1)\times SU(n)$ (see for example \cite{sym}). 
The black holes solution and the
entropy, both at the classical and quantum level, 
for the first set of models has been discussed in 
\cite{ka/sh}-\cite{be}. 
In this letter we will discuss a general black hole solution
related to the second set of model. The entropy for these models has
recently been discussed in \cite{sa}. 
For the case of $n=1,$ we get the axion-dilaton black hole
\cite{be/ka/or}.
The models which we will describe correspond to the most general quadratic
prepotential or more generally to the case where $M_I$ is proportional to 
$L^I$. For this  model the symplectic section is given by
\be
\pmatrix{L^I \cr M_I}= \pmatrix{L^I\cr\Sigma_{IJ} L^J}, 
\eq
with $\Sigma_{IJ} =\alpha_{IJ}-i\beta_{IJ}$, where
$\alpha_{IJ}$ and $\beta_{IJ}$ are arbitrary real symmetric matrices.

It is clear that the above section is related by a symplectic transformation 
to one in which 
$M_I=-i\beta_{IJ}L^I.$
This can be easily seen from the relation 
\be
\pmatrix{L^I\cr \Sigma_{IJ} L^J}=\pmatrix{L^I\cr 
(\alpha_{IJ}-i\beta_{IJ}) L^J}=\pmatrix{{1}&0\cr \alpha_{IJ}&1}
\pmatrix{L^I\cr -i\beta_{IJ} L^J}. 
\eq
and the fact that
\be
\pmatrix{{1}&0\cr \alpha_{IJ}&1}\in Sp(2n+2). 
\eq
Therefore, in the analysis of our model, one can, alternatively, start to 
find a solution for $\Sigma_{IJ}=-i\beta_{IJ}$ and then perform a symplectic 
transformation to obtain the solution for the more general $\Sigma$.

We shall first start with the discussion of the entropy of our model. 
Near the horizon, the 
extremization of the central charge
gives the following relations between the moduli fields and 
the charges \cite{FerKal1}
\be
\pmatrix{p\cr q}=Re\pmatrix{2i\bar ZL\cr 2i\bar ZM}.
\eq
In general this relation expresses the moduli fields 
$z^A={L^A\over L^0}$ in terms of the 
conserved electric and magnetic charges and topological data. 
Substituting these values back in 
$\vert Z\vert^2$ and
multiplying  by $\pi$ yields the Beckestein-Hawking entropy $\cal S$,
\be
{\cal S}=\pi\vert Z\vert^2_{hor}.
\eq
In matrix notation, the extremization equations
take the simple form
\be
i(Y-\bar Y)=p\qquad i({\Sigma}Y-\bar{\Sigma}\bar Y)=q,\qquad Y=\bar ZL.
\eq
This gives for our model
\be
Y={i}({\Sigma}-\bar{\Sigma})^{-1}(\bar{\Sigma}p-q),
\eq
and
\be
\vert Z\vert^2_{hor}={i}(p^t\bar{\Sigma}-q^t)({\Sigma}-\bar{\Sigma})^{-1}
(q-{\Sigma}p)
\eq
which can be rewritten as
\be
\vert Z\vert^2_{hor}=u^t{\cal V}u
\eq
where 
\be 
u=\pmatrix{q\cr p},\qquad {\cal V}={i}\pmatrix{(\bar\Sigma-{\Sigma})^{-1}
&({\Sigma}-\bar{\Sigma})^{-1}{\Sigma}\cr\bar{\Sigma}
({\Sigma}-\bar{\Sigma})^{-1}&
\bar{\Sigma}(\bar\Sigma-{\Sigma})^{-1}\Sigma}.
\eq
In terms of the components of ${\Sigma}=\alpha-i\beta$, the entropy 
can be expressed as
\be
{\cal S}=\pi\vert Z\vert^2={\pi\over2}\pmatrix{q^t&p^t}\pmatrix{\beta^{-1}
&-\beta^{-1}\alpha\cr -\alpha\beta^{-1}&\beta+\alpha\beta^{-1}\alpha}
\pmatrix{q\cr p}
\eq
For
\be
{\Sigma}=-i{\cal L},\qquad {\cal L}=\pmatrix{0&1\cr 1&0}
\eq
we get
\be
{\cal S}={\pi\over2}\pmatrix{q^t&p^t}\pmatrix{{\cal L}&0\cr 0&{\cal L}}
\pmatrix{q\cr p}
\eq
which are the entropies discussed for the cosets 
$SU(1,n)\over U(1)\times SU(n)$ in \cite{sa}.

It is clear that the value of the entropy for a general $\Sigma$ with
real and imaginary parts can be obtained from the expression of the entropy
of the model where the real part of $\Sigma$ is set to zero.
This is because the two formulation, as discussed earlier, are
connected via a symplectic transformation.  
The two formulations correspond to different embeddings of the
duality group of the theory into the symplectic group related by a 
symplectic similarity transformation. 
As an illustration, consider the classical model where 
${\Sigma}=-i\eta ,$ with $\eta=\hbox{diag}(1,-1,-1,\cdots,-1)$
for which the entropy is given by \cite{sa} 
\be
{\cal S}=\pi\vert Z\vert^2={\pi\over2}\pmatrix{q^t&p^t}
\pmatrix{\eta&0\cr 0&\eta}\pmatrix{q\cr p}
\label{en}\eq
The duality symmetry of this model is given by $SU(1,n)$. If we 
represent an element $M$ of $SU(1,n)$ as follows
\be
M=U+iV
\eq
Then the entropy in (\ref{en}) corresponds to a formulation 
(for which an $F$ function exists) with an  
embedding of 
$SU(1,n)$ into 
$Sp(2n+2)$ under which the quantum numbers transform by \cite{yaya}
\be
\pmatrix{q\cr p}\pmatrix{\eta U\eta&-\eta{V}\cr-{V}\eta&-{U}}\pmatrix{q\cr p}
={\cal R}\pmatrix{q\cr p}.\label{com}
\eq
Using the relations satisfied by $U$ and $V$
\be
U^t\eta U+V^t\eta V=\eta, \qquad U^t\eta V-V^t\eta U,
\eq
it is easy to see that the entropy (\ref{en}) is invariant under (\ref{com}).

Now consider for the model with $\Sigma=\alpha-i\eta$, with quantum numbers 
denoted by $(Q,P)$. 
The entropy in terms of $(Q, P)$ can be obtained from 
(\ref{en}) by noting that
\be
\pmatrix{q\cr p}=\pmatrix{1&-\alpha\cr 0&1}\pmatrix{Q\cr P}
\eq
Clearly the new entropy, given by
\be
{\cal S}=\pi\vert Z\vert^2={\pi\over2}\pmatrix{Q^t&P^t}\pmatrix{\eta
&-\eta\alpha\cr -\alpha\eta&\eta+\alpha\eta\alpha}
\pmatrix{Q\cr P}
\eq
is invariant under the following 
transformation
\be
\pmatrix{Q\cr P}\rightarrow \pmatrix{1&\alpha\cr 0&1}
{\cal R}\pmatrix{1&-\alpha\cr 0&1}\pmatrix{Q\cr P}={\cal R}'\pmatrix{Q\cr P}.
\eq
It is obvious that ${\cal R}'$ represents another embedding of $SU(1,n)$ into 
$Sp(2n+2)$.
Therefore the duality
symmetry does not get modified due to the correction $\alpha$ and depends 
only on the form of $\beta$.

We now turn to discuss static extreme black hole solution to our model.
The $N=2$ supergravity action includes one gravitational, $n$ vector and 
hyper multiplets. In what follows we will neglect the hyper multiplets,
assuming that these fields are constant. The bosonic $N=2$ action is
given by
\be 
S \sim \int
d^{4}x \sqrt{G} \{ R - 2 g_{A\bar{B}} \partial z^A \, \partial
\bar{z}^B + {1 \over 4 } (\Im {\cal N}_{IJ} F^I \cdot F^J + 
\Re {\cal N}_{IJ} F^{I}\cdot {^{\star}F^{J}})\} 
\eq 
The gauge field couplings  
are given in terms of the holomorphic
prepotential $F(X)$ (when it exists) by
\be 
{\cal N}_{IJ} =
\bar{F}_{IJ} + 2i {(\Im F_{IL}) (\Im F_{MJ}) X^{L} X^{M} 
\over(\Im F_{MN}) X^{M} X^{N}} 
\eq 
with $F_{I} = {\partial F(X)\over\partial X^{I}}$ and 
$F_{MN}={\partial^{2} F(X)\over\partial X^{M}\partial X^{N}},$ 
where the gauge field part 
$F^I \cdot F^J \equiv F^I_{\mu\nu} F^{J \,\mu \nu}$ and 
$I,J = 0,1,\cdots, n$. 

The most general form of an $N=2$ static black hole solution is given by 
\cite{tod}
\be
ds^2 =-e^{-2 U} dt^2 +  e^{2 U} d\vec{x} d\vec{x}, \qquad
\eq
and our solution is
\begin{equation} \label{solution}
\begin{array}{l} 
e^{2U} = \pmatrix{H^t&\tilde{H}^t}
{\cal V}\pmatrix{H\cr\tilde{H}},\\
F^I_{\mu\nu}=\epsilon_{\mu\nu\rho}\partial_\rho\tilde{H}^I\quad,\quad
G_{I\, \mu\nu} =\epsilon_{\mu\nu\rho}\partial_\rho (H_I)\quad,\quad
z^I = {Y^I \over Y^0}\\
Y=i(\Sigma-\bar\Sigma)^{-1}(\bar\Sigma\tilde H-H) \quad , \quad
G_{I\, \mu\nu}=\Re {\cal N}_{IJ} F^J_{\mu\nu}-
\Im {\cal N}_{IJ}{^{\star}F^{J}_{\mu\nu}}
\end{array}
\end{equation}
where
$H$ and $\tilde H$ are vectors with harmonic functions with components 
given by
\be
\tilde{H}^{I} =( h^{I} + {  p^{I} \over r} )\qquad,\qquad 
H_{I} =( h_{I} + { q_{I} \over r } )
\eq
and $h^{A}$, $h_{0}$ are constant and related to the scalar fields
at infinity.
The electric and magnetic charges are defined by integrals
over the gauge fields at spatial infinity
\be
q_I =\int_{S^2_{\infty}} G_{I\, \mu\nu},\qquad
p^I = \int_{S^2_{\infty}} F^I_{\mu\nu}.
\eq
To get the ADM mass we have to look on the asymptotic geometry. First, in 
order to have asymptotically a Minkowski space we have the constraint 
$e^{2U}\rightarrow 1$ for $r \rightarrow \infty$. This imposes the constraint
\be
\pmatrix{h^t&\tilde{h}^t}{\cal V}\pmatrix{h\cr\tilde{h}}=1.
\eq
Then using 
\be
e^{-2U} = 1 - {2 M \over r}+\cdots
\eq
we obtain the mass
\be
M={1\over2}\pmatrix{h^t&\tilde{h}^t}({\cal V}+{\cal V}^t)\pmatrix{q\cr {p}}
\eq
Clearly, $M^2$ should be equal to the central charge $|Z|^2$ 
evaluated at infinity. This will impose further constraints on the 
solutions. To see this, we rewrite the central charge in terms
of its real and imaginary parts as follows
\be
Z=Y^tq - Y^t\Sigma p={1\over2}
\pmatrix{H^t&\tilde{H}^t}({\cal V}+{\cal V}^t)\pmatrix{q\cr {p}}+{i\over2}
\pmatrix{h^t&\tilde{h}^t}\pmatrix{0&1\cr -1&0}\pmatrix{q\cr {p}}
\eq
It is then clear that 
\be
M^2 = |Z|^2_{\infty}.
\eq
provided that the imaginary part of the central charge is set to zero.
We should note that for our solution 
the imaginary part of the central charge is proportional to the 
chiral connection.

The vanishing of the imaginary part of the central charge indicates that
our solution should have an interpretation as a bound state. For $N=2$ black 
holes for supergravity models with cubic prepotential, we can take
as fundamental objects the 10d $D$-branes. In order to get a solution
with a non-singular horizon we have to compose four $D$-branes and as
result, the entropy in 4 dimensions behaves like $\sim \sqrt{q_0 p^1 p^2
p^3}$ (in the simplest case). For our solution given by eq. 
(\ref{solution}), the entropy is
given by a quadratic form in the charges. Hence,
one would expect that the bound state contains only two objects. In the 
simplest case we can regard our solution as an intersection
of two branes of  6d supergravity. These solutions are given by 0-,1- 
and 2-branes \cite{be/be/ja}
\begin{equation}
\begin{array}{l}
 ds^2_6 = {1 \over H} (-dt^2 + dz_p^2) + H\, dx_m^2 \quad , \quad 
 e^{-2 \phi} = H^{p-1}\\
 F = \left\{ \begin{array}{ll} dH \wedge dt \wedge \ldots  
 \wedge dz_p \quad , &  \mbox{for} \      p= 0,1 \\
    \ ^{*} dH \wedge dt \wedge \ldots  \wedge dz_p  \quad , & \mbox{for} 
    \ p=1,2
\end{array} \right. 
\end{array} 
\end{equation}
where $p= 0,1,2$ and $m= 1,\cdots,(5-p)$. In 6d (as in 10d) we have the
distinction between type IIA and IIB solutions. Here, the 0- and 2-branes
are IIA and the 1-brane is the self-dual IIB solution. All other brane
solutions are not asymptotically flat. In contrast to the 10d solutions,
the 6d branes depend on the harmonic function $H$ and not on $\sqrt{H}$
and thus all possible intersections give in 4d a general quadratic form
in the harmonic functions. These branes can of course be seen as
compactified intersections of 10d branes with identical charges. One can
however speculate about a different origin, namely as solutions of the
world volume theory of 5-branes. The 0-brane can be seen as an open
string ending on the $D$-5-brane. The self-dual string is known to appear
as solution of the $M$-5-brane, it is an open membrane ending on the
5-brane. Finally, the 2-brane is an open 3-brane ending on the $F$-5-brane. 
Thus, it
is tempting to speculate that our 4d black hole solutions are
compactified solutions of the 5-brane world volume theory, which appears as
$D$-5-brane as well as 
in $M$- and  $F$-theory.

In conlusion, we have discussed static black
hole solution of $N=2$ supergravity with a general quadratic
prepotential. In the first part we discussed the solution near the
horizon. Following the idea of extremization of the central charge, 
we derived the Bekenstein-Hawking  entropy and confirmed known results 
\cite{sa}. 
In the second part, we investigated a general solution for 
general moduli. This solution differs from the solution for a cubic
prepotential by i) that the black does not contain a square root of a
quartic form but is defined in terms of a quadratic form of harmonic
functions; ii) it can be solved explicitly for all charges, electric 
and magnetic, as well as general complex moduli. It turned out, 
that for the general 
charge configuration we have an additional constraint for the moduli, which
ensures that $M=\vert Z\vert_\infty$. This condition is the vanishing 
of the chiral connection, which in turn is necessary to have a static 
solution. For the known solutions 
related to a cubic prepotential,
this constraint is fulfilled identically \cite{be}. A non-vanishing
chiral connection should be related to a non-vanishing Taub-NUT charge or
angular momentum. The solution discussed in this paper is therefore a
good starting point to discuss these questions for $N=2$ black holes and
we will leave it for a forthcoming paper.
\vskip 1cm

\noindent {\bf Acknowledgements}  
\medskip \newline
The work is  supported by the DFG.
\renewcommand{\arraystretch}{1}

\end{document}